\documentclass[conference]{IEEEtran}

\ifCLASSINFOpdf
   \usepackage[pdftex]{graphicx}
\else
   \usepackage[dvips]{graphicx}
\fi

\usepackage{cite}
\usepackage[cmex10]{amsmath}
\usepackage{algorithmic}
\usepackage{array}
\usepackage{mdwmath}
\usepackage{mdwtab}
\usepackage{eqparbox}
\usepackage{stfloats}

\begin{document}
%
\title{Turbo DPSK in Bi-directional Relaying}
%
\author{Weijun Zeng\IEEEauthorrefmark{1},\IEEEauthorblockN{Xiaofu Wu\IEEEauthorrefmark{2}, and
Zhen Yang\IEEEauthorrefmark{2}} \\
\IEEEauthorblockA{\IEEEauthorrefmark{1}
PLA University of Science and Technology, Nanjing 210007, CHINA\\ Email: zwj3103@126.com}
\\ \IEEEauthorblockA{\IEEEauthorrefmark{2} Nanjing University of Posts and Telecommunications, Nanjing 210003, CHINA\\ Email: xfuwu@ieee.org, and yangz@njupt.edu.cn}}


\maketitle

\begin{abstract}
In this paper, iterative differential phase-shift keying (DPSK) demodulation and channel decoding scheme is investigated for the Joint Channel decoding and physical layer Network Coding (JCNC) approach in two-way relaying systems. The Bahl, Cocke, Jelinek, and Raviv (BCJR) algorithm for both coherent and noncoherent detection is derived for soft-in soft-out decoding of DPSK signalling over the two-user multiple-access channel with Rayleigh fading. Then, we propose a pragmatic approach with the JCNC scheme for iteratively exploiting the extrinsic information of the outer code. With coherent detection, we show that DPSK can be well concatenated with simple convolutional codes to achieve excellent coding gain just like in traditional point-to-point communication scenarios. The proposed noncoherent detection, which essentially requires that the channel keeps constant over two consecutive symbols, can work without explicit channel estimation. Simulation results show that the iterative processing converges very fast and most of the coding gain is obtained within two iterations.
\end{abstract}

\section{Introduction}

\begin{figure}[t]
\begin{center}
\includegraphics[scale=0.4]{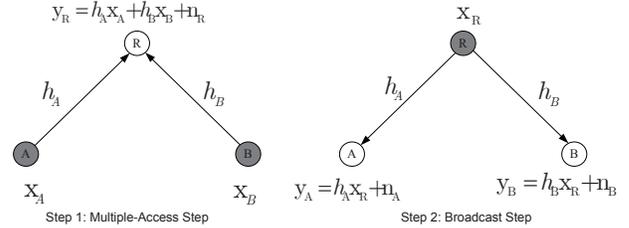}
\end{center}
\caption{Two-way relaying systems with DNF}
\label{fig:sys}
\end{figure}

\IEEEPARstart{W}{ireless} network coding has recently received a lot of attention for disseminating information over wireless networks \cite{KoetterAlg,Chou}. Indeed, the gain is very impressive for the special bi-directional relaying scenarios with two-way or multi-way traffic as addressed in \cite{Popovski}. For bi-directional relaying, two sources A and B want to exchange information with each other by the help of a relay node R as shown in Fig. \ref{fig:sys}.  Traditionally, this can be achieved via four steps. Recently, it was recognized that only two steps are essentially required with the employment of the powerful idea of physical-layer network coding (PNC) \cite{Zhang_PLNC}. This denoise-and-forward (DNF) approach has been shown to perform better than various other approaches \cite{Zhang_PLNC}.

For bi-directional relaying with PNC,  it is assumed that communication takes place in two phases - a multiple-access phase and a broadcast phase as shown in Fig. \ref{fig:sys}. In the first phase, the two source nodes send signals simultaneously to the relay.
In the second phase, the relay processes the superimposed signal of the simultaneous packets and
maps them to a network-coded (XOR) packet for broadcast back to the source nodes. Then, both sources can retrieve their own information  as they know completely what they have sent. Compared with the traditional relay system, PNC doubles the throughput of
the two-way relay channel.

For the DNF approach, most of the existing works focus on the coherent detection, where the perfect channel-state information (CSI) is assumed to be available at the relay \cite{Popovski,ZhangJSAC,LangLDPC,WubbenBPSK}. However, the CSI at the relay can be much more difficult to obtain compared to the traditional point-to-point channel as two source signals can interfere each other at the relay. To overcome this difficulty, it is interesting to explore the idea of noncoherent deteciton as did in the point-to-point communication \cite{Hoeher}. For noncoherent detection, differentially-encoded signalling is often employed as no explicit carrier recovery is required. Indeed, maximum likelihood (ML) detectors for both amplify-and-forward (AF) and DNF protocols are derived in \cite{CuiDiff} for differentially-encoded signalling over bi-directional relaying channels without CSIs assumed at the receivers.

To be more practical, channel coding should be employed to further improve the reliability of the system. In \cite{ZhangPNC,ZhangJSAC}, a combined channel coding and PNC scheme, namely, the Joint Channel decoding and physical layer Network Coding (JCNC), have been introduced. It was recognized that with the same linear channel code at both source nodes, the XOR of both source codewords is still  a valid codeword. Thus, the received signal can be decoded to the XOR of the two source messages at the relay without changing the decoding algorithm.

In this paper, the JCNC approach with differential signalling and channel coding is considered as shown in Fig. \ref{fig:relaydec}. It is clear that the performance bottleneck lies in the multiple-access phase where two sources signals can interfere each other and the coherent detection may become impractical. Hence, we focus on the design of powerful differential coding schemes, along with the noncoherent detection at the relay. Compared to the counterpart of the traditional point-to-point communication, it is interesting to investigate the performance of well-designed serially-concatenated codes with DPSK as the inner code over the multiple-access channel encountered at the relay. Furthermore, iterative demodulation and decoding schemes, without resort to channel estimation, are extremely welcome in the multiple-access channel. Essentially, it requires to develop a soft-in soft-output noncoherent demodulation algorithm for two superimposed DPSK signalling impaired by the unknown fading but constant over the period of at least two consecutive symbols.

The rest of the paper is organized as follows. The system model is introduced in section II. In section III, the soft-in soft-out algorithms under both coherent and noncoherent detections are derived and a pragmatic approach for iterative demodulation and decoding is proposed. Simulation results are given in section IV, and Section-V concludes the paper.

\section{System Model}

\begin{figure}[htbp]
\begin{center}
\includegraphics[scale=0.40]{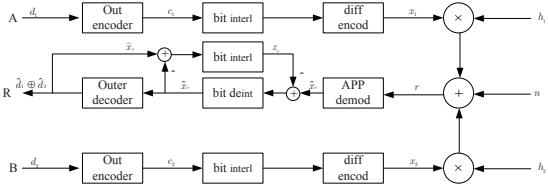}
\end{center}
\caption{System model}
\label{fig:relaydec}
\end{figure}

Throughout this paper, we focus on the multiple-access (MA) phase since the broadcast (BC) phase is degraded to the conventional point-to-point communication. The system model is shown in Fig. \ref{fig:relaydec}.

\subsection{Transmitter}

During the MA phase, consider two binary information bit sequences of length $L$, ${\bf{d}}_i  = \{ d_i (k)\}$, $i = 1,2$ to be transmitted by the sources A ($i=1$) and B ($i=2$), where each $d_i (k)$ $(k = 0,\cdots,L-1)$ can take values from $\{ 0,1\}$. At each source node, the binary information sequence is first encoded by a rate-$L/N$ linear outer coder and output the coded sequence of length $N$. Then, the coded sequence is interleaved and  further modulated using differential PSK signaling. For the DNF approach, it is assumed that both sources adopt the same encoder, interleaver and modulator.

Let ${\bf{c}}_i  = \{ c_i(0),c_i(1),\cdots,c_i (N-1)\}$ denotes the coded sequence (vector) for source $i$. Let $x_i (k) = E_s^{1/2} u_i (k)$  denote the complex baseband modulated signal at the $k$th time epoch,, where $E_s$
is the energy of the signal, $u_i(k) = e^{j\phi _i (k)}$ and $\phi _i (k) \in \{ 0,\pi \}$. For DPSK modulation, the information is carried in the phase difference of two consecutive modulated signals,
 \begin{equation}
\begin{array}{l}
$$\phi _i (k) = \phi _i (k - 1) + \Delta \phi _i (k)$$
 \end{array}
\end{equation}
where $\Delta \phi _i (k) = \pi$ corresponds to the input coded bit $c_i (k) = 0$ and $\Delta \phi _i (k) = 0$ corresponds to $
c_i (k) = {\rm{1}}$. It follows that
 \begin{equation}
    u_i (k) = u_i(k - 1) \cdot (2c_i(k)-1).
\end{equation}
\subsection{Channel Model}

Assume that the signals between source and relay are undergoing the correlated Relayeigh flat-fading channel, the received signal at the relay for the $k$ time epoch  can be given by
\begin{equation}
    r(k) = h_1 (k)x_1 (k) + h_2 (k)x_2 (k) + n(k),
\end{equation}
where the fading coefficients $h_1 (k)$ and $h_2 (k)$ are zero-mean complex Gaussian random variables with variances $2\sigma _1 ^2$, $2\sigma _2 ^2$, respectively. Here, the fading process is assumed to be slow such that the coefficient $h_i$ $(i=1,2)$ can be considered to remain constant during at lease two consecutive symbol intervals. In other words, $h_i(k) \approx h_i(k - 1)$ for all $k$. $n(k)$ is a complex Gaussian random variable with mean zero and variance $2\delta ^2$. The collection of the received signal vector of length $N+1$ is denoted by $\mathbf{r}_0^N =(r_0,\cdots,r_N)$, where $r_0$ corresponds to the initial reference symbol for differential-encoded PSK signalling.

\subsection{The DNF Approach}

The denoise-and-forward approach is first proposed in \cite{PopovskiVTC}. For this approach, the relay employs a denoising function based on an adaptive network coding to map the received signal vector ${\bf{r}}$ into a quantized signal vector ${\bf{c}}_{\rm{R}}$ for the broadcast phase. In this paper, the standard XOR operation is employed as the denoising mapper, namely, ${\bf{c}}_{\rm{R}}  = {\bf{c}}_1 \oplus {\bf{c}}_2$, where $\oplus$ denotes a bit-wise XOR operation.

Due to the employment of the same linear outer code and interleaver at both sources, the decoding can be performed jointly to obtain the XOR of
the information from the two sources. Thus the relay does not have to decode the information from the two sources separately.

\section{Iterative Demodulation and Decoding for the DNF Approach}

\subsection{Coherent BCJR-Type Algorithm}
For the BCJR-type algorithm, it is essential to provide a trellis description of two differential PSK signals superimposed at the relay. With coherent detection, one means that both fading processes $h_i(k), i=1,2$ are known to the relay. Let us define, at time epoch $k$, the state $s_k$ as
\begin{equation}
    s_k = \left(u_1(k-1), u_2(k-1) \right).
\end{equation}
and the branch metric function as
\setlength{\arraycolsep}{0pt}
\begin{eqnarray}
\label{eq:sd}
   \gamma_k && (s_k,\mathbf{c}_{12}(k)) \propto \Pr(\mathbf{c}_{12}(k)) \nonumber   \\
    &&  \cdot \exp\left(-\frac{\left|r(k)- h_1(k)x_1(k)-h_2(k)x_2(k)\right|^2}{2\delta^2}\right),
\end{eqnarray}
\setlength{\arraycolsep}{5pt}
where $\mathbf{c}_{12}(k) = (c_1(k),c_2(k))$ denotes the couple of coded bits at time epoch $k$ for both sources .

\begin{figure}[htbp]
\begin{center}
\includegraphics[scale=0.40]{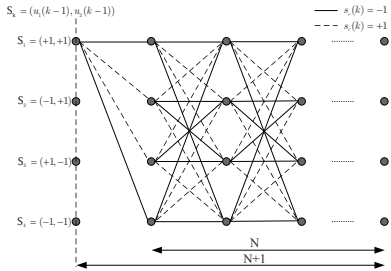}
\end{center}
\caption{Trellis diagram of 2-DPSK in the relay}
\label{fig:treDPSK}
\end{figure}

This trellis description can be found in Fig. \ref{fig:treDPSK}, which has $2^2$ states and $2^2$ branches per state for differentially-encoded $2$-PSK.

The BCJR algorithm is characterized by the following forward
and backward recursions:
\begin{eqnarray}
\label{eq:fd}
   \alpha_{k+1}(s_{k+1})=\sum_{\mathbf{c}_{12}(k)} \sum_{s_k} \mathcal{T}(\mathbf{c}_{12}(k),s_k,s_{k+1}) \nonumber \\
    \cdot \alpha_k(s_k) \gamma_k(s_k,\mathbf{c}_{12}(k)),
\end{eqnarray}
where $\mathcal{T}(\mathbf{c}_{12}(k), s_k, s_{k+1})$ is the trellis indicator function, which is equal to 1 if $\mathbf{c}_{12}(k), s_k$, and $s_{k+1}$
satisfy the trellis constraint and 0 otherwise.
\begin{eqnarray}
\label{eq:fd}
   \beta_{k}(s_k)=\sum_{\mathbf{c}_{12}(k)} \sum_{s_{k+1}} \mathcal{T}(\mathbf{c}_{12}(k),s_k,s_{k+1}) \nonumber \\
   \cdot \beta_{k+1}(s_{k+1}) \gamma_k(s_k,\mathbf{c}_{12}(k)).
\end{eqnarray}
Then, the joint APPs $\Pr\left(\mathbf{c}_{12}(k)|\mathbf{r}_0^{N}\right)$ can be calculated as
\setlength{\arraycolsep}{0.0em}
\begin{eqnarray}
\label{eq:llr}
    \Pr &&\left(\mathbf{c}_{12}(k)|\mathbf{r}_0^{N}\right)  \nonumber \\
    && =\sum_{s_{k+1}}\mathcal{T}(\mathbf{c}_{12}(k),s_{k+1}) \alpha_{k+1}(s_{k+1}) \beta_{k+1}(s_{k+1}),
\end{eqnarray}
where the indicator function $\mathcal{T}(\mathbf{c}_{12}(k),s_{k+1})$ is equal to 1 if $s_{k+1}$
is compatible with $\mathbf{c}_{12}(k)$ and 0 otherwise.

Finally, the log-likelihood ratio at time epoch $k$ is computed as
\begin{eqnarray}
  \label{eq:13}
  L_r(k) &=& \log\frac{\Pr(c_R(k)=1|\mathbf{r}_0^{N})}{\Pr(c_R(k)=0|\mathbf{r}_0^{N})} \nonumber \\
      &=&  \log\frac{\sum_{c_1(k) \oplus c_2(k)=1}\Pr(\mathbf{c}_{12}(k)|\mathbf{r}_0^{N})}{\sum_{c_1(k) \oplus c_2(k)=0}\Pr(\mathbf{c}_{12}(k)|\mathbf{r}_0^{N})}.
\end{eqnarray}

\subsection{Noncoherent BCJR-Type Algorithm}

For the noncoherent detection, we assume that both fading processes $h_i(k), i=1,2$ are unknown to the relay but keep constant over at least two symbols. Hence, one can still construct the trellis as did in the coherent approach. However, the branch metric should be modified as
\setlength{\arraycolsep}{0pt}
\begin{eqnarray}
\label{eq:sd}
   \gamma_k && (s_k,\mathbf{c}_{12}(k)) \propto \Pr(\mathbf{c}_{12}(k)) \nonumber   \\
    &&  \cdot p\left(r(k),r(k - 1)|\mathbf{x}_{12} (k),\mathbf{x}_{12} (k - 1)\right),
\end{eqnarray}
\setlength{\arraycolsep}{5pt}
where $\mathbf{x}_{12}(k)=(x_1(k),x_2(k))$.

Hence, it is the task of noncoherent algorithm to calculate the conditional probability density function (pdf) $p\left(r(k),r(k - 1)|\mathbf{x}_{12} (k),\mathbf{x}_{12} (k - 1)\right)$.
The conditional pdf can be written as
\begin{eqnarray}
 p\left(r(k),r(k - 1)|\mathbf{x}_{12} (k),\mathbf{x}_{12} (k - 1)\right) \nonumber \\
  = p(r(k)|\mathbf{x}_{12}(k),\mathbf{x}_{12}(k-1),r(k - 1)) \nonumber \\
 {\rm{   }} \cdot p(r(k - 1)|\mathbf{x}_{12}(k),\mathbf{x}_{12}(k-1)) \nonumber \\
  = p(r(k)|\mathbf{x}_{12}(k),\mathbf{x}_{12}(k-1),r(k - 1)) \nonumber \\
 {\rm{   }} \cdot p(r(k - 1)|\mathbf{x}_{12}(k-1)).
\end{eqnarray}

For the Rayleigh fading process, it is clear that $r(k - 1)$ is the sum of three zero-mean complex Gaussian variables given the transmitted symbols $x_1 (k - 1)$
, $x_2 (k - 1)$. Hence, it is also a zero-mean complex Gaussian variable. The variance of  $r(k - 1)$ given $\mathbf{x}_{12} (k - 1)$ can be calculated as
\setlength{\arraycolsep}{0pt}
 \begin{eqnarray}
 D &[& r(k - 1)| \mathbf{x}_{12}(k - 1)]  \nonumber \\
  &=& D[h_1 (k - 1))|x_1 (k - 1)|^2]  \nonumber \\
  &&+ D[h_2 (k - 1))|x_2 (k - 1)|^2]  + D[n(k - 1)] \nonumber \\
 &=& 2\sigma _1 ^2 E_s  + 2\sigma _2 ^2 E_s  + 2\delta ^2
\end{eqnarray}
\setlength{\arraycolsep}{5pt}
where $E_s$ is the energy of the signal.

Accordingly, the conditional pdf can be written as
\begin{eqnarray}
 p(r(k - 1)|\mathbf{x}_{12}(k - 1)) = \frac{1}{{2\pi \sigma _r^2 }}\exp \left( - \frac{{|r(k - 1)|^2 }}{{2\sigma _r^2 }}\right).
\end{eqnarray}

Note that $r(k)$ and $r(k - 1)$ are jointly-Gaussian distributed given $\mathbf{x}_{12}(k), \mathbf{x}_{12}(k-1)$. Hence, $p\left(r(k)|\mathbf{x}_{12}(k - 1),\mathbf{x}_{12}(k),r(k - 1)\right)$ is Gaussian distributed \cite{PapouRand}. Let $m_r$, $\delta _r^2$ be the mean and the variance of $r(k)$ given $\mathbf{x}_{12}(k)$, $\mathbf{x}_{12}(k-1)$, $r(k - 1)$, respectively. It is straightforward to show that
\setlength{\arraycolsep}{0pt}
\begin{eqnarray}
 m_r &=& E\left[r(k)|\mathbf{x}_{12}(k),\mathbf{x}_{12}(k - 1),r(k - 1)\right] \nonumber \\
   &=& \frac{{E[r^* (k)\cdot r(k - 1)|\mathbf{x}_{12}(k), \mathbf{x}_{12}(k - 1)]}}{{E[r^* (k - 1)\cdot r(k - 1)|\mathbf{x}_{12}(k),\mathbf{x}_{12}(k - 1)]}} \cdot r(k - 1) \nonumber \\
   &=& \frac{{2x_1 (k)x_1 ^* (k - 1)\sigma _1 ^2  + 2x_2 (k)x_2 ^* (k - 1)\sigma _2 ^2 }}{{\sigma _r^2 }}\cdot r(k - 1), \nonumber \\
\end{eqnarray}
and
\begin{eqnarray}
 \delta _r^2  &=& E[r^* (k)\cdot r(k)|\mathbf{x}_{12}(k),\mathbf{x}_{12}(k - 1)] \nonumber \\
  &-&  \frac{{\left\{E[r^* (k)\cdot r(k - 1)|\mathbf{x}_{12}(k),\mathbf{x}_{12}(k - 1)]\right\}^2 }}{{E[r^* (k - 1) \cdot r(k - 1)|\mathbf{x}_{12}(k),\mathbf{x}_{12}(k - 1)]}} \nonumber \\
  &=& \sigma _r^2  - \frac{{4E_s ^2 \sigma _1 ^4  + 4E_s ^2 \sigma _2 ^4 }}{{\sigma _r^2 }} \nonumber \\
  &+& \frac{{8x_1 (k)x_1 ^* (k - 1)x_2 (k)x_2 ^* (k - 1)\sigma _2 ^2 \sigma _1 ^2 }}{{\sigma _r^2 }}.
\end{eqnarray}

\subsection{Iterative Demodulation and Decoding}

Once the soft LLRs referred to the network-coded codeword ${\bf{c}}_{\rm{R}}$ are computed, the relay node can perform channel decoding for obtaining the pairwise XOR of the two source bits if both sources A and B assume the same linear channel code. This JCNC approach \cite{ZhangJSAC} has an obvious advantage in complexity.

Although this JCNC approach can be efficiently implemented, it does result in performance loss due to the use of $\Pr(c_1(k)\oplus c_2(k)|\mathbf{r}_0^N)$ while the joint probabilities $\Pr(c_1(k),c_2(k)|\mathbf{r}_0^N)$ is not fully used. In \cite{WubbenBPSK}, a generalized sum-product algorithm (G-SPA) over the Galois field $GF(2^2)$ was proposed for low-density parity-check (LDPC) coded BPSK systems, which can work directly with the joint probabilities and a significant gain was observed compared to the JCNC approach.

From the viewpoint of complexity, the JCNC has significant advantage over the G-SPA. Hence, we focus on the JCNC in this paper.

With serially-concatenated DPSK signalling, the coding gain cannot be fully achieved without employing the iterative demodulation and decoding. In what follows, we propose a pragmatic approach for exploiting the soft extrinsic information provided by the outer decoder when the JCNC scheme is employed. Let us denote the soft extrinsic information by $L_e\left(c_R(k)\right)$, which is produced by the outer soft-in soft-out decoder.  As there is a no one-to-one correspondence between $c_R(k)$ and $\mathbf{c}_{12}(k)$, we have to resort to the following pragmatic method, namely,
\setlength{\arraycolsep}{0pt}
\begin{eqnarray}
\label{eq:sd}
   \Pr(c_1(k)=0, c_2(k)=0) &\approx& \Pr(c_1(k)=1, c_2(k)=1) \nonumber \\
                           &\approx& \frac{1}{2} \Pr(c_1(k) \oplus c_2(k)=0) \nonumber \\
                           & \propto & \exp\left(-\frac{1}{2} L_e\left(c_R(k)\right)\right), \\
   \Pr(c_1(k)=0, c_2(k)=1) &\approx& \Pr(c_1(k)=1, c_2(k)=0) \nonumber \\
                           & \propto & \exp\left(\frac{1}{2} L_e\left(c_R(k)\right)\right).
\end{eqnarray}
\setlength{\arraycolsep}{5pt}
Besides of this pragmatic use of extrinsic information, iterative demodulation and decoding is the same as that of \cite{Hoeher}. Hence, we omit the details.

\section{Simulation Results}

The bottleneck of the bi-directional relaying system lies in the processing capability of the relay node. For various joint channel coding and PNC schemes, it is the duty of the relay to reproduce the XOR of both source codewords. Hence, we mainly focus on the performance of the XOR codeword $\mathbf{c}_R$. The performance is closely related to the energy per bit and the received noise variance.

Two outer codes are considered. The first one is a (3,6)-regular Mackay-Neal LDPC code with codewords of length $N=1008$\cite{MacKay}, and the second one is a rate-1/2 convolutional code with generator $(23,35)_8$ with the information bit length of $N/2=504$. The iterative sum-product algorithm (SPA) is employed for decoding of LDPC codes and the maximum number of decoding iterations is set to 20. For decoding of convolutional codes, the optimal BCJR algorithm is employed.

\begin{figure}[t]
\begin{center}
\includegraphics[scale=0.90]{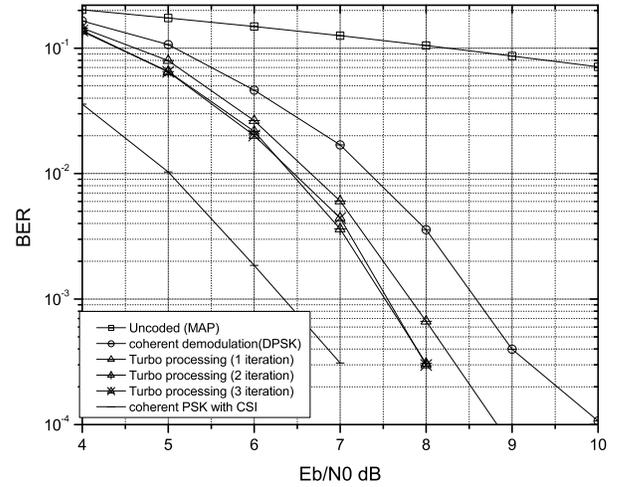}
\end{center}
\caption{Differentially encoded BPSK: Performance for Rayleigh fading assuming perfect channel estimation with the outer LDPC code.}
\label{fig:cohLDPC}
\end{figure}

\begin{figure}[t]
\begin{center}
\includegraphics[scale=0.90]{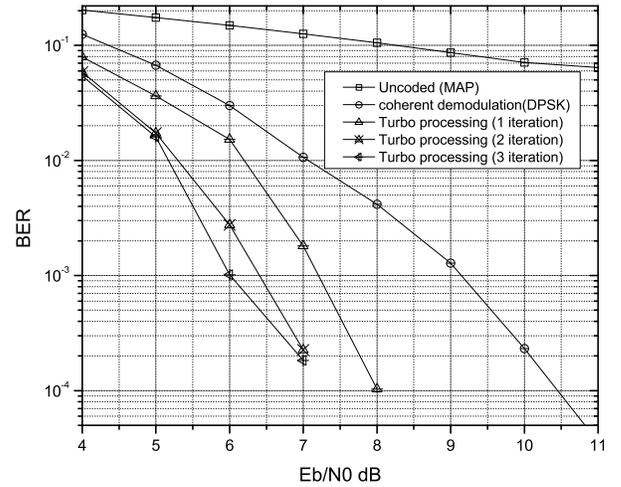}
\end{center}
\caption{Differentially encoded BPSK: Performance for Rayleigh fading assuming perfect channel estimation with the outer convolutional code.}
\label{fig:cohConv}
\end{figure}

\begin{figure}[t]
\begin{center}
\includegraphics[scale=0.90]{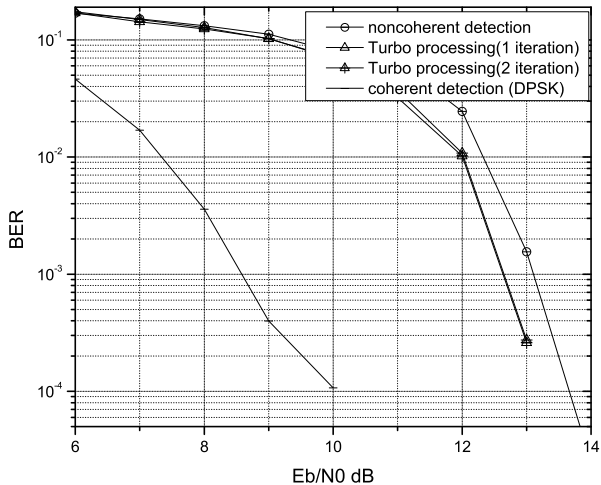}
\end{center}
\caption{Differentially encoded BPSK: Performance for Rayleigh fading without channel estimation with the outer LDPC code.}
\label{fig:nonLDPC}
\end{figure}

\begin{figure}[t]
\begin{center}
\includegraphics[scale=0.90]{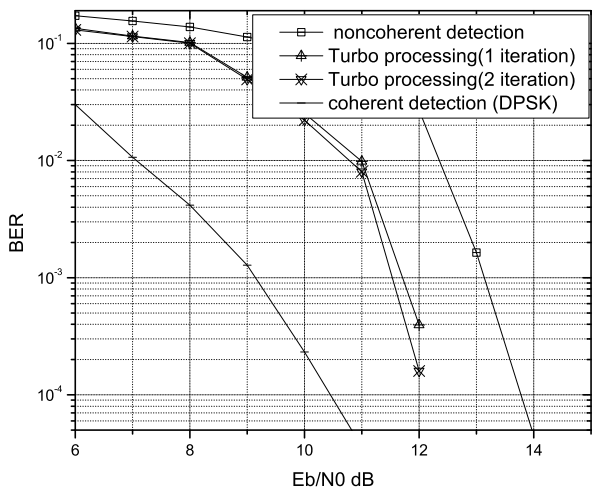}
\end{center}
\caption{Differentially encoded BPSK: Performance for Rayleigh fading without channel estimation with the outer convolutional code.}
\label{fig:nonConv}
\end{figure}

For Rayleigh fading channels, the Jakes model is adopted with a normalized fade rate of $f_d T_s$.

Firstly, we consider the coherent detection by assuming perfect channel estimation. For the fading rate of $f_d T_s  = 0.03$, the BER performance is shown in Fig. \ref{fig:cohLDPC} and Fig. \ref{fig:cohConv} for LDPC and convolutional codes, respectively.  As shown, the proposed pragmatic iterative processing does work and  the iterative processing gain is about $1.5$ dB for the LDPC code and 3 dB for the convolutional code at $10^{ - 4}$. It is shown that the convergence is very fast and  3 iterations are enough. With more than 3 iterations, the performance improvement is minor. It is also shown that the simple outer convolutional code can achieve better coding gain compared to the Mackey-Neal LDPC code when serially-concatenated with DPSK.

Secondly, we consider noncoherent detection without channel estimation. With the number of states restricted to 4, the noncoherent BCJT-type algorithm can be efficiently implemented. For the same fading rate of $f_d T_s  = 0.03$, the BER performance is shown in Fig. \ref{fig:nonLDPC} and Fig. \ref{fig:nonConv}. At the BER of $10^{ - 4}$, the iterative processing gain is about 1 dB for the LDPC code and 2 dB for the convolutional code. As shown, the convergence of the noncoherent approach is even faster compared to the coherent approach and it is enough to consider 2 iterations. Compared to the ideal coherent approach, there is about 5 dB loss for the noncoherent approach in signal-to-noise ratios at the BER of $10^{-4}$, both with iterative processing.

\section{Conclusion}
In this paper, we have investigated the performance of serially-interleaved concatenated codes under iterative processing for the DNF approach in bi-directional relaying. Similar to the traditional point-to-point communication, the simple convolutional codes can be serially concatenated with differential PSK modulation for achieving the significant coding gain in bi-directional relaying scenarios. It should be emphasized that the coding gain can be achieved with a pragmatic iterative processing approach, which can be efficiently implemented.

\section*{Acknowledgment}
This work was supported in part by the National Science Foundation of China under Grants 60972060, 61032004. The work of X. Wu was also supported by the National Key S\&T Project under Grant 2010ZX03003-003-01. The work of Yang was also supported by the National Science and Technology Major Project under grant 2010ZX0 3003-003-02, and by the National Basic Research Program of China (973 Program) under grant 2011CB302903.

\end{document}